\RequirePackage{fix-cm}
\documentclass[smallextended]{svjour3}       
\smartqed  

\usepackage{subfigure}
\usepackage{graphicx}
\graphicspath{ {./images/} }
\usepackage{threeparttable}
\usepackage{amsfonts}
\usepackage{cite}
\graphicspath{ {Figure/} }
\usepackage{comment}
\usepackage{amsmath}
\usepackage{bm}

\begin{document}

\title{Aliasing Instabilities in the Numerical Evolution of the Einstein Field Equations
}


\author{C. Meringolo         \and
        S. Servidio 
}


\institute{C. Meringolo \at
              Dipartimento di Fisica, Universit\`a della Calabria, I-87036 Cosenza, Italy \\
              \email{claudiomeringolo@unical.it}           
           \and
}

\date{Received: date / Accepted: date}

\maketitle
 
\begin{abstract}
The Einstein field equations of gravitation are characterized by cross-scale, high-order nonlinear terms, representing a challenge for numerical modeling. In an exact spectral decomposition, high-order nonlinearities correspond to a convolution that numerically might lead to aliasing instabilities. We present a  study of this problem, in vacuum conditions, based on the $3+1$ Baumgarte-Shibata-Shapiro-Nakamura (BSSN) formalism. We inspect the emergence of numerical artifacts, in a variety of conditions, by using the Spectral-FIltered Numerical Gravity codE (\texttt{SFINGE}) - a pseudo-spectral algorithm, based on a classical (Cartesian) Fourier decomposition. By monitoring the highest $k-$modes of the dynamical fields, we identify the culprits of the aliasing and propose procedures that cure such instabilities, based on the suppression of the aliased wavelengths. This simple algorithm, together with appropriate treatment of the boundary conditions, can be applied to a variety of gravitational problems, including those related to massive objects dynamics.
 

\end{abstract}

\section{Introduction}
In the last decades, numerical relativity has become one of the most successful approaches to understand extreme astrophysical situations \cite{diener2006accurate, baker2006binary, Scheel2006solving, herrmann2006binary, Schnetter2004evolutions}. To this end, supercomputers are often employed to study black holes, neutron stars, gravitational waves, and many other phenomena governed by Einstein's theory of gravitation. A very active field of research in numerical relativity is the simulation of relativistic binaries and their associated gravitational waves. Building a numerical spacetime is a difficult task since the equations consist of a set of multidimensional, nonlinear, partial differential equations (PDEs). Because of the intrinsic nonlinearity and complexity of the Einstein field equations, several works have been devoted to the study of appropriate numerical methods \cite{laguna2002numerical, witek2011stability, Bernuzzi2010constraint, calabrese2002stability, Zlochower2005accurate, Jansen2006numerical,Pretorius2006simulation}.

The common approach to the Einstein equations relies on slicing the four-dimensional spacetime into three-dimensional spacelike hypersurfaces. From the mathematical point of view, this procedure allows treating the system as a Cauchy problem with constraints. From a practical point of view, it reduces to a decomposition of the spacetime into ``space'' + ``time'', so that one manipulates only time-varying tensor fields in the three-dimensional space. Here we deal with the Baumgarte-Shapiro-Shibata-Nakamura (BSSN) decomposition \cite{Baumgarte1998numerical, Shibata1995evolution, Beyer2004well, Brown2008bssn, Sarbach2002hyperbolicity}, by proposing a very accurate pseudo-spectral strategy, in vacuum conditions, based on classical Fast Fourier Transforms (FFTs). 

Spectral methods are very performing techniques for numerical relativity. In this regard, two well-known spectral algorithms are the SpEC \cite{Scheel2006solving} and the SGRID code \cite{Tichy2009long}, that represent a clear example of highly performing codes. These important studies suggest that the spectral decomposition might be both accurate and stable. In the spectral approach, the nonlinear terms become convolutions in the complex (Fourier) space \cite{Boyd2001chebyshev}, and numerical problems might arise from the so-called ``aliasing instabilities'' \cite{Orszag1971elimination, Gottlieb1977numerical}. These instabilities generally start from the higher Fourier-modes and propagate back to smaller $k-$vectors (larger scales) during the simulation. This transfer of information is a purely numerical artifact, with no physical meaning and the numerical simulation generally loses accuracy, with a consequent blow-up of the solution. The dealiasing procedure plays a crucial role, as we shall see in this paper.

We proceed here within the path of spectral methods, concentrating on a simple algorithm, namely the Spectral-FIltered Numerical Gravity codE (SFINGE) -- a pseudo-spectral algorithm, based on a classical, Cartesian-Fourier decomposition. We propose adequate numerical strategies in order to prevent numerical errors and instabilities. The algorithm, proposed by Meringolo et al \cite{Meringolo2021spectral}, has been successfully tested in a variety of conditions. The SFINGE code makes use of {\it pseudo-spectral} techniques \cite{Orszag1972comparison, Orszag1979spectral}, where one continuously switches back-and-forth between the real (physical) and the complex (Fourier) space. The approach is quite simple: in order to avoid convolutions, products are computed in the physical space while derivatives are computed in the spectral space. Classical spectral and pseudo-spectral approximations are widely used for nonlinear problems that involve strong couplings.  Both techniques have the advantage that each field and all its derivatives are represented at the collocation points. This leads to very precise solutions, with an error on the computation of derivatives that is on the order of the machine truncation error \cite{Canuto2012spectral, Dutykh2016brief, Grandclement2009spectral, Orszag1971elimination}. With respect to previous algorithms of numerical relativity, our code keeps the simplicity of the Cartesian domain, with periodic boundary conditions. Alternative boundaries have been implemented, with the same technique, by using absorbing spherical dissipation, as described in Meringolo et al. \cite{Meringolo2021spectral}, and as briefly summarized in the present work. Finally, the code is able to manage extreme situations such as the head-on collision of black holes, by using appropriate absorbing boundaries and filtering procedures. 
 
In this work, we investigate the role of artificial instabilities, by monitoring the behavior of the isotropic spectra of the dynamical variables and, in particular, the evolution of the highest modes in the Fourier space - the modes close to the Nyquist wavenumber. We show that these modes rapidly grow up during the simulation, leading to the inaccuracy of the results. We propose a way to prevent these numerical problems, based on the dealiasing of high-order products. In practice, we remove these aliasing instabilities and improve the numerical simulations by constructing dealiasing filters \cite{Hou2007computing}. In this way, the BSSN code becomes stable and very accurate, with an excellent agreement with analytical solutions - whenever they are available. Such a technique can be used also in compact object simulations.

The work is organized as follows. In Section \ref{sec:num} we present the governing BSSN equations together with the numerical code, while in Section \ref{sec:test} we perform a series of gravitational testbeds, controlling the numerical instabilities and proposing the strategy to suppress such problems. Finally, in Section \ref{discuss}, we present our conclusions.

\section{The numerical model}
\label{sec:num}
The BSSN decomposition is a formalism based on the work by Baumgarte and Shapiro \cite{Baumgarte1998numerical} and by Shibata and Nakamura \cite{Shibata1995evolution}. This formulation starts from a conformal rescaling of the physical metric
\begin{equation}
    \gamma_{ij} = \psi^{4} \widetilde{\gamma}_{ij}, 
    \label{deco}
\end{equation}
where $\psi$ is a conformal factor. We follow the approach of Campanelli \textit{et al.} \cite{Campanelli2006accurate}, where this factor is written as $\chi = \psi^{-4}$. This choice has been demonstrated to be a better alternative when considering singular spacetimes for which $\psi$ typically has a $r^{-1}$ singularity, while $\chi$ is a $C^4$ function at the singularity. By following this particular choice, the Ricci tensor can be separated into two contribution, $R_{ij} = \widetilde{R}_{ij} + R^{\chi}_{ij}$, where 
\begin{eqnarray}
\nonumber
\widetilde{R}_{ij}= - \frac{1}{2} \widetilde{\gamma}^{lm} \partial_l \partial_m   
    \widetilde{\gamma}_{ij}
    + \widetilde{\gamma}_{k(i} \partial_{j)} \widetilde{\Gamma}^k 
    + \widetilde{\Gamma}^k  \widetilde{\Gamma}_{(ij)k}  
     + \widetilde{\gamma}^{lm}  \Big( 2 \widetilde{\Gamma}^k_{l(i}  \widetilde{\Gamma}_{j)km} 
    + \widetilde{\Gamma}^k_{im}  \widetilde{\Gamma}_{klj} \Big)
\end{eqnarray}
is the Ricci tensor related to the conformal metric and 
\begin{eqnarray} 
\begin{split}
\nonumber
  R_{ij}^\chi = \frac{1}{2 \chi} \Bigg\{ 
    \bigg[
    \partial_i \partial_j \chi
   - \frac{\partial_i \chi \, \partial_j \chi}{2 \chi}
   - \widetilde{\Gamma}^k_{ij} \partial_k \chi  \bigg]
   +  \widetilde{\gamma}_{ij} \bigg[
    \widetilde{\gamma}^{lm} \bigg( 
    \partial_l \partial_m \chi 
    - \frac{3 \, \partial_l \chi \, \partial_m \chi}{2 \chi}   \bigg)
    - \widetilde{\Gamma}^k \partial_k \chi
    \bigg]
    \Bigg\}
\end{split}    
\end{eqnarray}
is the part that depends on the conformal factor $\chi$. In the BSSN approach, the extrinsic curvature $K_{ij}$ is divided in two independents variables, the trace $K$ and its trace-free parts $A_{ij}$. The last one is subjected to the same conformal transformation described by equation (\ref{deco}), i.e. $\widetilde{A}_{ij} = \psi^{-4} \big( K_{ij} - \frac{1}{3} \gamma_{ij} K \big)$. Finally, a new field is introduced, namely the contracted Christoffel symbols associated with the conformal metric  $\widetilde{\Gamma}^i = \widetilde{\gamma}^{jk} \widetilde{\Gamma}^i_{jk}$. 

With the above change of variables, the system of BSSN equations reads:
\begin{equation}
        \partial_0 \widetilde{\gamma}_{ij} = - 2 \alpha \widetilde{A}_{ij},
        \label{evo1} 
\end{equation}
\begin{equation}
        \partial_0 \chi = \frac{2}{3}\chi \alpha K ,
        \label{evo2} 
\end{equation}
\begin{equation}
        \partial_0   K = - D^2 \alpha + \alpha \Big( \widetilde{A}_{lm}\widetilde{A}^{lm} +  \frac{1}{3} K^2 \Big),        
        \label{evo3}
\end{equation}
\begin{equation}
        \partial_0  \widetilde{A}_{ij} =
        \chi \Big[ - D_i D_j \alpha + \alpha R_{ij}
        \Big]^{TF}+
        \alpha \Big(K \widetilde{A}_{ij} - 2 \widetilde{A}_{ik} \widetilde{A}^{k}_{j}\Big),
        \label{evo4} 
\end{equation}
\begin{equation}
\begin{split}
        \partial_t \widetilde{\Gamma}^i =  \widetilde{\gamma}^{lm} \partial_l \partial_m \beta^i + \frac{1}{3} \widetilde{\gamma}^{il} \partial_l \partial_m \beta^m  +  \beta^k \partial_k \widetilde{\Gamma}^i - \widetilde{\Gamma}^k \partial_k \beta^i + \frac{2}{3} \widetilde{\Gamma}^i \partial_k \beta^k   \\ 
         -  2 \widetilde{A}^{ik} \partial_k \alpha  
         + \alpha \Big(  2\widetilde{\Gamma}^i_{lm} \widetilde{A}^{lm}
         - \frac{3}{ \chi}\widetilde{A}^{ik} \partial_k \chi 
         - \frac{4}{3} \widetilde{\gamma}^{ik} \partial_k K  \Big).
    \label{evo5}
    \end{split}
\end{equation}
In the above equations note that  $\partial_0 = \partial_t - \mathcal{L}_\beta$, where $\mathcal{L}_\beta$ is the Lie derivative with respect to the shift $\beta^k$, $\alpha$ is the lapse, $D_i$ is the covariant derivative associated with the physical three-metric $\gamma_{ij}$ and $D^2 = \gamma^{ij} D_i D_j$. As usual, ``$TF$'' indicates the trace-free part of a tensor (i.e. for a generic tensor $T_{ij}$ one has $[T_{ij}]^{TF}= T_{ij} - \frac{1}{3} \gamma_{ij} \gamma^{lm}T_{lm}$).  We now need the evolution equation for the lapse and the shift, choosing therefore the slicing conditions. In the framework of the Bona-Massó formalism \cite{Bona1989einstein, Bona1995new,Baumgarte1998numerical}, we choose
\begin{equation}
\partial_0 \alpha = -\alpha^2 f(\alpha) K, 
\label{slicing1}
\end{equation}
\begin{equation}
\partial_0 \beta^i = \frac{3}{4} B^i ,
\label{slicing2} 
\end{equation}
\begin{equation}
\partial_0  B^i = \partial_t \widetilde{\Gamma}^i - \beta^j \partial_j  \widetilde{\Gamma}^i - \eta B^i. 
\label{slicing3}
\end{equation}
Here $\eta$ is a positive constant and is set to $\eta = 2.8$ \cite{Alcubierre2003towards}, and the factor $3/4$ is somewhat arbitrary but leads to good numerical results. 

In addition to the evolution equations, as discussed in \cite{Zlochower2005accurate, Akbarian2015black, Alekseenko2004constraint, Gundlach2006well, Brown2009covariant}, the BSSN variables must satisfy the vacuum constraints
\begin{equation}
    \label{ham}
    \mathcal{H}= R - \widetilde{A}_{lm} \widetilde{A}^{lm} + \frac{2}{3} K^2=0,
\end{equation}
\begin{equation}
    \label{mom}
    \mathcal{M}^i=\partial_k \widetilde{A}^{ik} + \widetilde{\Gamma}^i_{lm} \widetilde{A}^{lm}- \frac{3}{2 \chi } \widetilde{A}^{ik} \partial_k \chi - \frac{2}{3} \widetilde{\gamma}^{ik} \partial_k K =0,
\end{equation}
\begin{equation}
    \mathcal{G}^i=  \widetilde{\Gamma}^i + \partial_j \widetilde{\gamma}^{ij}=0,
\end{equation}
\begin{equation}
    \widetilde{\gamma} - 1 = 0,
        \label{cons1} 
\end{equation}
\begin{equation}
     \widetilde{A} = 0,
    \label{cons2} 
\end{equation}
where $\widetilde{\gamma}$ is the determinant of the conformal metric, and $\widetilde{A}$ is the trace of $\widetilde{A}_{ij}$. During the numerical simulations, we enforce the algebraic constraints in equations (\ref{cons1}) and (\ref{cons2}). The remaining constraints, $\mathcal{H}$, $\mathcal{M}^i$ and $\mathcal{G}^i$ are not actively enforced and are used to monitor the accuracy of the numerical solutions \cite{Alcubierre2003towards}.  In all the above equations, $\widetilde{\Gamma}^i$ is replaced by $-\partial_j \widetilde{\gamma}^{ij}$, wherever it is not differentiated.

\subsection{The SFINGE code}
The numerical procedure is described in \cite{Meringolo2021spectral} and has been tested against all classical testbeds. The spatial derivatives are computed in Cartesian geometry, via standard FFTs. For each BSSN field $f(x^{\mu})\equiv f({\bm x}, t)$, the equations have been discretized on a equally-spaced lattice of $N_x\times N_y \times N_z$ mesh points. At the collocation points, $f({\bm x}, t)= \sum _{{\bf k}} \widetilde{f}_{\bf k}(t) ~ \exp(i  {\bm k}\cdot{\bm x})$, with ${\bm x}$ being the positions of the nodes, $\widetilde{f}_{\bf k}(t) \in \mathbb{Z}$ the Fourier coefficients and the wavevectors ${\bm k}= (k^x, k^y, k^z)$. 
Along each box side $L_0$, the wavevector is $k=2\pi m/L_0$, with $m=0, \pm 1, \pm 2, \dots \pm N_k$, where  $N_k=N/2$ the Nyquist mode. The code is based on a parallel architecture and makes use of MPI directives and FFTW libraries \cite{Cooley1969fast}.

In the spectral space, the nonlinear terms become a convolution and there are several transform-based techniques for evaluating it efficiently \cite{Canuto2012spectral, Boyle2007testing,  Cooley1965algorithm}.  Numerical problems might arise because of the so-called aliasing instabilities \cite{Hossain1992computing}: any product among fields creates higher $k$'s, causing an alias when these new modes become larger than the maximum size $N_k$ of sampled Fourier modes. A Fourier mode with a wavenumber out of the size range is aliased to another wavenumber in the domain and, in a time-evolving system, it creates growing numerical problems. The importance of eliminating such pathology (aliasing error) has been studied since pioneering works by Orszag et al. \cite{Orszag1971elimination}.

In equations with low-order nonlinearities, such as incompressible Navier-Stokes, the dealiasing technique is nothing but making the maximum number of Fourier modes $k^*$ sufficiently smaller than the size $N_k$. For quadratic nonlinearities, it is only necessary to filter the top one-third of modes. If the cutoff wavenumber is $N_k$ (equal to $N/2$ in one dimension) and if only modes with $|k|< k^* \equiv 2/3 N_k$ can be excited. In this case, the aliasing is completely removed. For systems with higher-order products, or whenever the nonlinearity is not well-defined, there are complications. Because of the high nonlinearity of the BSSN equations, the aliasing effect is much more pronounced and the transfer to larger $k$-vectors is more efficient. If one considers, for example, the evolution equation for $\widetilde{A}_{ij}$ (Eq.~\ref{evo4}), there are products of the type
\begin{eqnarray}
\nonumber
    \partial_t \widetilde{A}_{ij} \sim \dots + \alpha  \chi^3 \gamma_{ik} \gamma^{mk} \gamma_{mj} K^2 + \dots.
\end{eqnarray}
The above quantity corresponds to a challenging convolution in the Fourier space. Moreover, other complications come from the contravariant tensors, which involve further products. These high-order products immediately lead to large harmonics and therefore spectral ambiguity. 

We use a strategy to mitigate the above problems, based on an analogy with compressible hydrodynamics. We inspect the role of the aliasing truncation \cite{Frisch2008hyperviscosity, Shu2005numerical}, by using a filtering technique.  The method is very simple to implement: for any product one smoothly sets ${\widetilde Q}_k = 0$ for $k>k^\star$.  In practice, for each representation one has
\begin{equation}
    f({\bm x},\it t) =  \sum _{{\bm k}} \widetilde{f}_{\bm k}(t) e^{i k_j x^j}\Phi_{k^*}({\bm k}),
\label{fltr}
\end{equation} 
where the spectral anti-aliasing filter given by
\begin{equation}
	\Phi_{k^*}({\bm k})= e^{-a\left[\frac{|{\bm k}|}{k^*}\right]^2 }.
	\label{filteralias1}
\end{equation}
Here $a$ is a free parameter (here $a=20$) that gives the sharpness of the filter and $k^*$ is again the cutoff. Different values of $k^\star$ have been chosen, depending on the difficulty of the simulation and on the initial data.

For the time integration we adopt a second-order Runge-Kutta method, with a  time-step can be changed during the evolution, which simply corresponds to an \textit{adaptive time-refinement}  technique \cite{Meringolo2021spectral}.  For the timestep control we estimate a global characteristic time as $\mathcal{T}_f = \delta f /{\sqrt{ \Big \langle \left( \frac{\partial f}{\partial t}\right)^2} \Big \rangle}$,  where $\delta f=\sqrt{\langle \left(f -\langle f \rangle\right)^2 \rangle}$ is the $rms$ of the field and $\langle \dots \rangle$ indicates a volume average. The quantity can be interpreted as a typical dynamical time of $f$, evaluated as the ratio between the level of fluctuations and its typical time-derivative. Overall, the method consists of a self-adjustment of the Runge-Kutta time-step $\Delta t$, such that $\Delta t < C~\min_f \{ \mathcal{T}_f\}$, where $C$ can be regarded as the Courant number (we set $C=1/4$). 

Whenever we want to model non-periodic problems, such as black hole dynamics, we introduce an implicit, viscous strategy to absorb boundary disturbances, typical of the numerical tests that have a small violation of the periodicity. We implemented an implicit hyperviscous boundary (IHB) method. Loosely speaking, we use Cartesian geometries in order to simulate spherical domains and dissipate the regions close to the corners, with a radial envelope \cite{Schneider2005decaying, Dobler2006magnetic, Servidio2007compressible}. Hyperviscoscous terms of the type $\nu_n \nabla^n f$ (where $\nu_n$ is a numerical coefficient) are able to dissipate quickly ripples and numerical artifacts. For any dynamical BSSN variable, we advance in time the solution $f^{\{ideal\}}({\bf x}, t)$ by using the classical second-order Runge-Kutta method. Simultaneously, if the IHB is set on, we evolve in time an accompanying, twin-field $f^{\{H\}}({\bf x}, t)$ which obeys to the same BSSN equation but is also subject to hyperviscous dissipation, being therefore highly damped. We advance in time this hyperviscous field $f^{\{H\}}({\bf x}, t)$ by using a Crank--Nicolson, semi-implicit method (CN).  With this procedure, the internal ideal region, advanced via the Runge-Kutta method, is matched with the outer diffusive layer, integrated by the implicit CN scheme. Overall, the price to pay for the IHB is that we have to double the time of integration, slowing down the computation. On the other hand, the method gives clear benefits on the stability and the goodness of the solutions, as we shall see later, in subsection~\ref{Headon}.

\begin{table}
\caption{\label{table2} Table of simulations. From left to right: the run number, the initial condition, number of points, $k^*$ of the filter, the filter status, the RSC status, the IBH status, and finally, the stability of the simulation. Here ''\checkmark'' stands for on, and ''$\times$'' for  off. } 
{\def\arraystretch{1.2}\tabcolsep=4.0pt
\begin{tabular}{@{}lllllllll}
\hline\noalign{\smallskip}
 RUN & IC type             & $ N $   &  $k^*$   & Filter & RSC & IHB & Stable \\
\noalign{\smallskip}\hline\noalign{\smallskip}
1   & Gauge wave           &  $128$   & $\infty$ & $\times$   & $\times$ & $\times$  & no \\
2   & Gauge wave           &  $128$   & $N/3$    & \checkmark & $\times$  & $\times$  & yes \\ 
3   & Robust stability     &  $128$   & $\infty$    & $\times$   & $\times$ & $\times$  & no \\
4   & Robust stability     &  $128$   & $N/3$    & \checkmark & $\times$ & $\times$ & yes \\
5   & Gowdy wave           &  $64$   & $\infty$    & $\times$   & $\times$ & $\times$   & no \\
6   & Gowdy wave           &  $64$   & $N/4$    & \checkmark & \checkmark & $\times$ & yes \\
7   & Head on collision    &  $256^3$ & $\infty$    & $\times$   & $\times$  & $\times$  & no \\ 
8   & Head on collision    &  $256^3$ & $N/2$    & \checkmark & $\times$ & $\times$  & no \\ 
9   & Head on collision    &  $256^3$ & $N/2$    & \checkmark & \checkmark  & \checkmark  & yes \\ 
\noalign{\smallskip}\hline
\end{tabular} }
\end{table}

As the main diagnostic of the work, we compute the power spectra of all the BSSN variables and monitor them in time in order to follow and understand the development of instabilities. In the next section, we simulate a variety of gravitational situations via the above SFINGE algorithm, based on the filtering technique of Eq.s~(\ref{fltr})-(\ref{filteralias1}), highlighting the strategy to suppress the aliasing.

\section{Numerical tests}
\label{sec:test}

We present simulations of gravitational testbeds  \cite{Alcubierre2003towards, Dumbser2018conformal}, among the standard ``apples with apples'' tests. For each initial data, we follow the evolution of the spectral power of the fields, highlighting the hints of the aliasing instabilities, that usually manifest as an exponential pile-up of energy at small scales. We check the accuracy of the simulation (1) by matching the numerical result with the analytical one (where it is possible), (2) by inspecting the BSSN constraints, and (3) via the computation of power spectra. A list of simulations is reported in Table \ref{table2}.

\begin{figure}[t]
\hspace{-10pt}
\includegraphics[height=87mm,width=119mm]{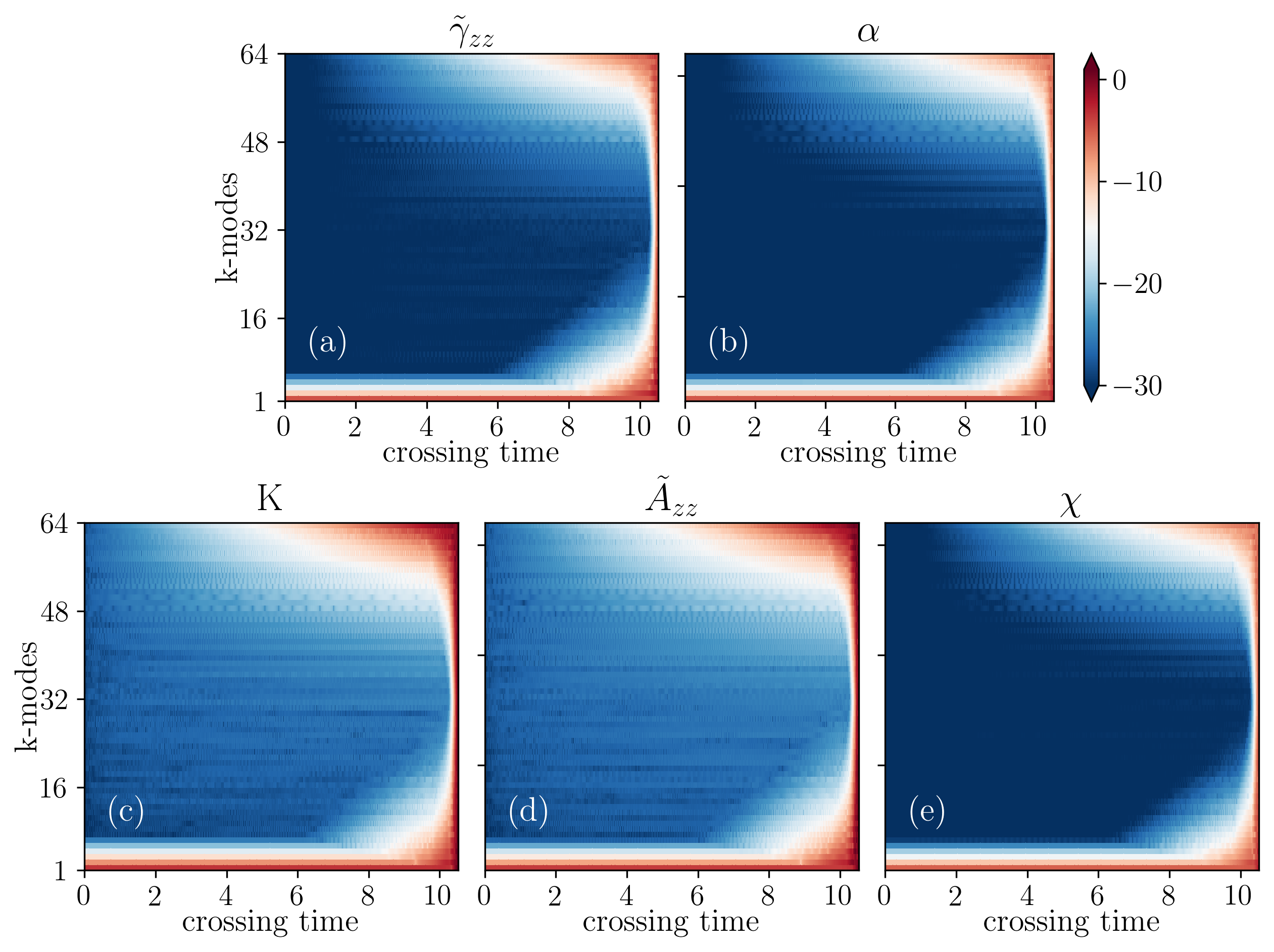}
\caption[...]
{\footnotesize{}Energy spectra of the main BSSN dynamical variables, as a function of time and $k$, for the small amplitude gauge wave test (RUN$_1$). No anti-aliasing filter has been used and the code become unstable at $t\sim 10.4$. From panel $(a)$ to $(e)$ we show $\widetilde{\gamma}_{zz}$, $\alpha$, $K$, $\widetilde{A}_{zz}$ and $\chi$, respectively.}
\label{gauge6}
\end{figure}

\subsection{Gauge wave}
The gauge wave test is based on a transformation of the Minkowski spacetime, with 
\begin{equation}
    ds^2 = -H(z,t)\, dt^2 \,+dx^2 +dy^2+H(z,t)\,dz^2, 
\label{wavetest1}
\end{equation}
where $H(z,t) = 1-A~\sin\big[2 \pi (z-t)\big]$ describes a sinusoidal modulation of amplitude $A$, propagating along the $z$-axis, with $A<1$. Since derivatives are zero in the $x$ and $y$ directions, the problem is essentially 1D. The metric in equation (\ref{wavetest1}) implies $\beta^i=0$, and $K_{zz}=-\partial_t \gamma_{zz} / 2 \alpha$. For the spatial metric one gets
\begin{eqnarray}
\nonumber
    \gamma_{ij}=
    \left( 
    \begin{matrix}
     1~~~ & 0~~~ & 0 \cr
     0~~~ & 1~~~ & 0 \cr
     0~~~ & 0~~~ & 1-A\sin[2 \pi (z-t)] \cr
     \end{matrix} 
     \right),
\end{eqnarray}
while the only nontrivial component of the extrinsic curvature is
\begin{eqnarray}
\nonumber
    K_{zz}=-A\pi\frac{ \cos[2 \pi (z-t)]}{\sqrt{1-A~\sin[2 \pi (z-t)]}}.
\end{eqnarray}
The above fields satisfy the initial data constraints in (\ref{ham})-(\ref{mom}).

\begin{figure}[t]
\hspace{-5pt}
\includegraphics[height=37mm,width=119mm]{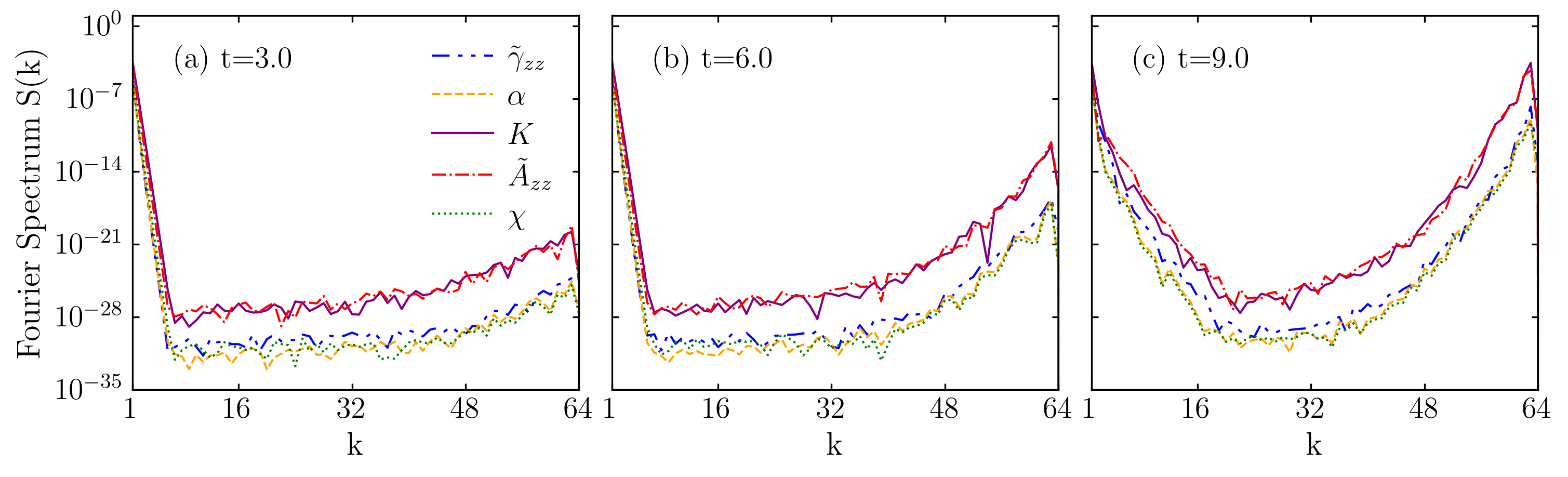}
\caption[...]
{\footnotesize{}Spectra of $\widetilde{\gamma}_{zz}, \alpha, K, \widetilde{A}_{zz}$ and $ \chi$, for RUN$_1$, at different times of the simulation.}
\label{gauge3}
\end{figure} 

We chose a spatial domain $x \in [ 0, 1 ]$ with a number of meshes $N_z = 128$. The Runge-Kutta time-step is $\Delta t = 10^{-3}$ and, according to the literature, we use the harmonic slicing in Eq.~(\ref{slicing1}), without shift evolution ($\beta^k = \partial_t \beta^k = 0$) \cite{Alcubierre2003towards}. The amplitude of the perturbation is $A=10^{-2}$. For the first run, RUN$_1$, no anti-aliasing filter ($k^*=\infty$) and no adaptive time refinement (RSC) has been used. The code crashes very quickly, at $10.4$ crossing times.

In order to better understand where the numerical instabilities arise from, we monitor the Fourier spectra of the main BSSN variables. In figure \ref{gauge6}, we report the Fourier spectrum for the BSSN variables, as a function of both time and wavenumber $k$. In particular, we show  $\widetilde{\gamma}_{zz}$, $\alpha$, $K$, $\widetilde{A}_{zz}$, and $\chi$, until the code crashes. Notice that a logarithmic scale has been used. From the above figure one can appreciate that the first hint of instability seems to come from the extrinsic curvature (the conformal trace-free part $\widetilde{A}_{zz}$ (panel $(d)$) and its trace $K$ (panel $(c)$). Then these instabilities gradually contaminate all the other dynamical variables until the code becomes unstable and blows up, at about $t \sim 10.4$.

In figure \ref{gauge3} we show the energy spectra of the same dynamical variables, at $t=3.0$, $6.0$ and $9.0$. Note that nonphysical ``energy'' contaminates first the higher Fourier modes, and then diffuses back to the whole spectrum. The major contribution comes from the extrinsic curvature.

\begin{figure}[t]
\includegraphics[height=83mm,width=119mm]{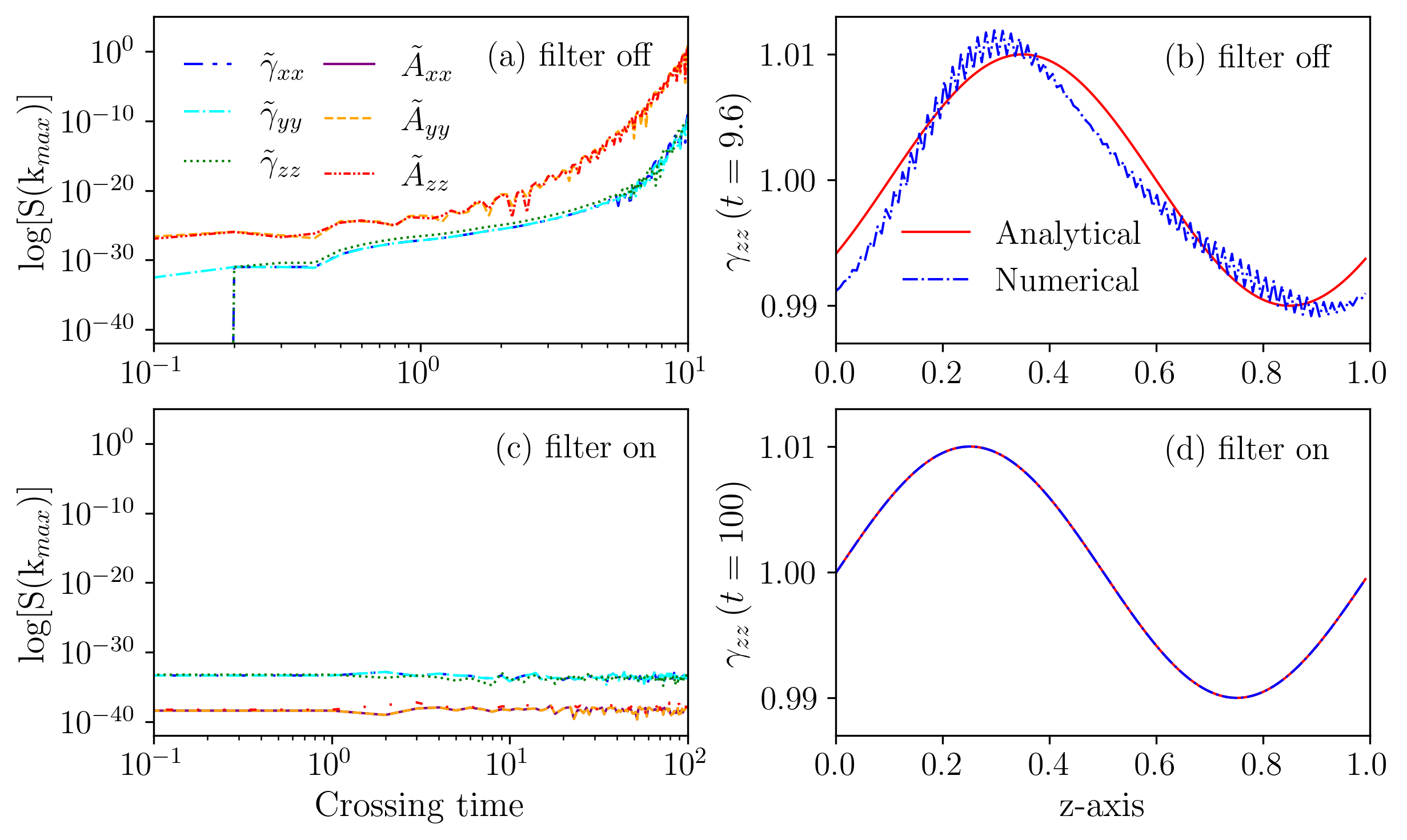}
\caption[...]
{\footnotesize{}Top row: small amplitude gauge wave test,  without filters (RUN$_1$). In panel $(a)$ we show the evolution in time of the $k=k_{max}$ Fourier mode of the diagonal elements for the conformal $\widetilde{\gamma}_{ij}$ and $\widetilde{A}_{ij}$. The aliasing instability leads to an exponential growth of such modes. In $(b)$ we report a comparison between the physical metric $\gamma_{zz}$ at $t=9.6$ and the exact solution of the wave-form. Bottom row: same for the filtered RUN$_2$, where the solution becomes very stable and accurate.}
\label{gauge1}
\end{figure}

It is interesting now to follow in time the modes. In figure \ref{gauge1}-(a) we show the evolution of $k_{max}\sim N/2$, for the diagonal elements of the conformal metric and the conformal trace-free part of the curvature. The exponential growth in time is evident. This {\it  spectral check} shows clearly that aliasing errors arise first in the trace-free part of the extrinsic curvature and then in the metric components. The consequences of such instabilities are visible in figure \ref{gauge1} (b),  where we show a section of the physical three-metric $\gamma_{zz}$ at $t=9.6$, compared with the analytical solution. It is interesting to note not only the appearance of wiggles but also a strong deformation of the largest wavelength (initial) mode.

\begin{figure}[t]
\includegraphics[height=90mm,width=119mm]{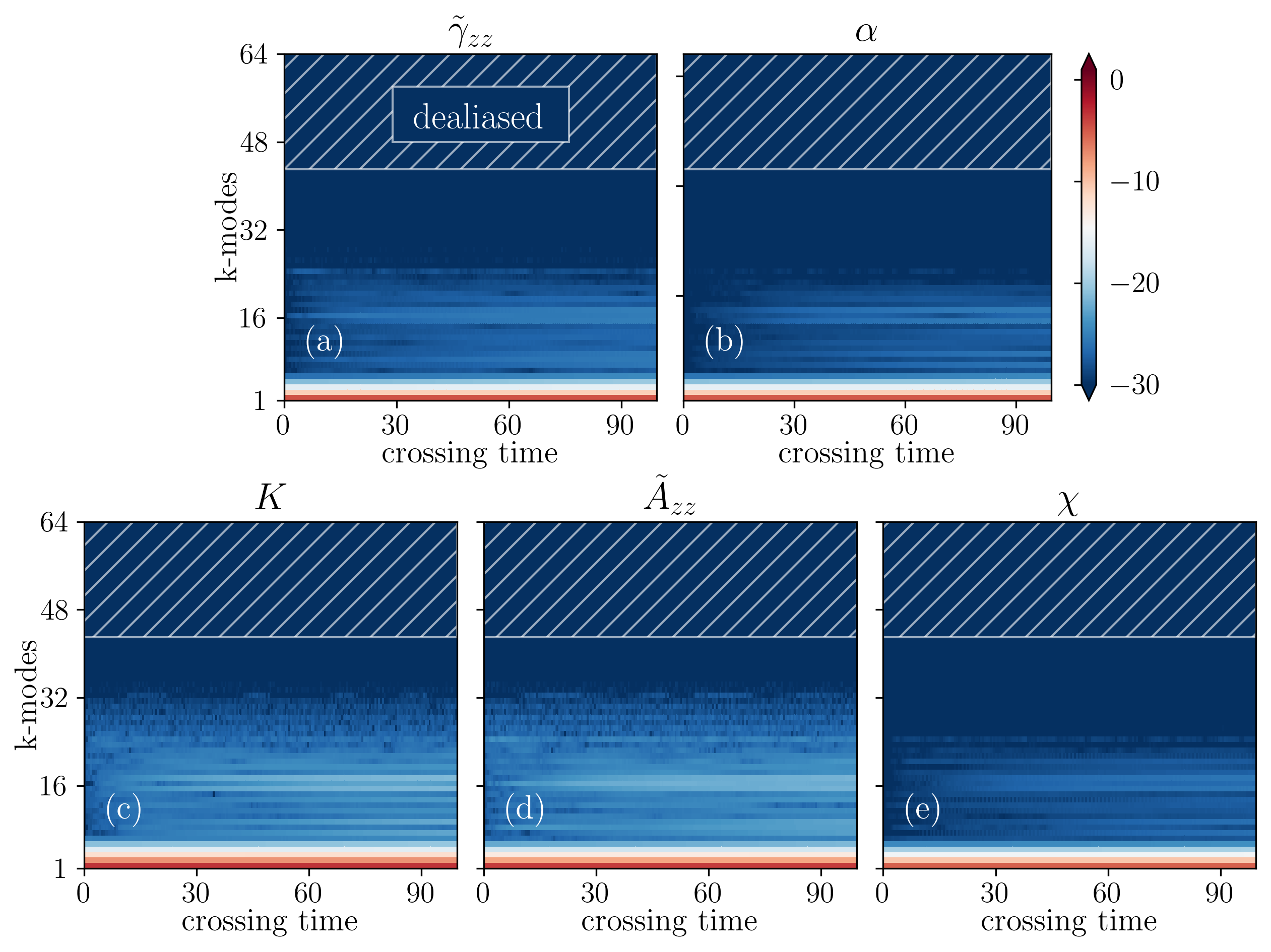}
\caption[...]
{\footnotesize{} Energy spectra of the main BSSN dynamical variables as a function of $k$ and $t$, for RUN$_2$. A smooth filter with $k^* = N_z/3$ has been used. The code remains stable for long time.}
\label{gauge6_t100}
\end{figure}

In order to avoid these instabilities and stabilize the algorithm, in the second run (RUN$_2$) we use our smooth anti-aliasing filter described by equations (\ref{fltr})-(\ref{filteralias1}), using $k^* = N_z/3$ (the other parameters are the same as the previous run.) In figure \ref{gauge1}-(c) we show the evolution of the $k=k_{max}$ Fourier-mode. The filter suppresses the non-physical energy in all the highest harmonics and the simulation remains stable for long time. The advantage is evident from panel (d), where we show a section of the $\gamma_{zz}$ component at $t=100$ crossing time, compared with the exact solution. Finally, in figure \ref{gauge6_t100}, we show the evolution in time of the spectra of the main BSSN variables, for the RUN$_2$. One can see that, in contrast with the previous test, the smooth filter suppresses higher Fourier modes and the code remains very stable.

\subsection{Robust stability test}
\begin{figure}
\includegraphics[height=77mm,width=119mm]{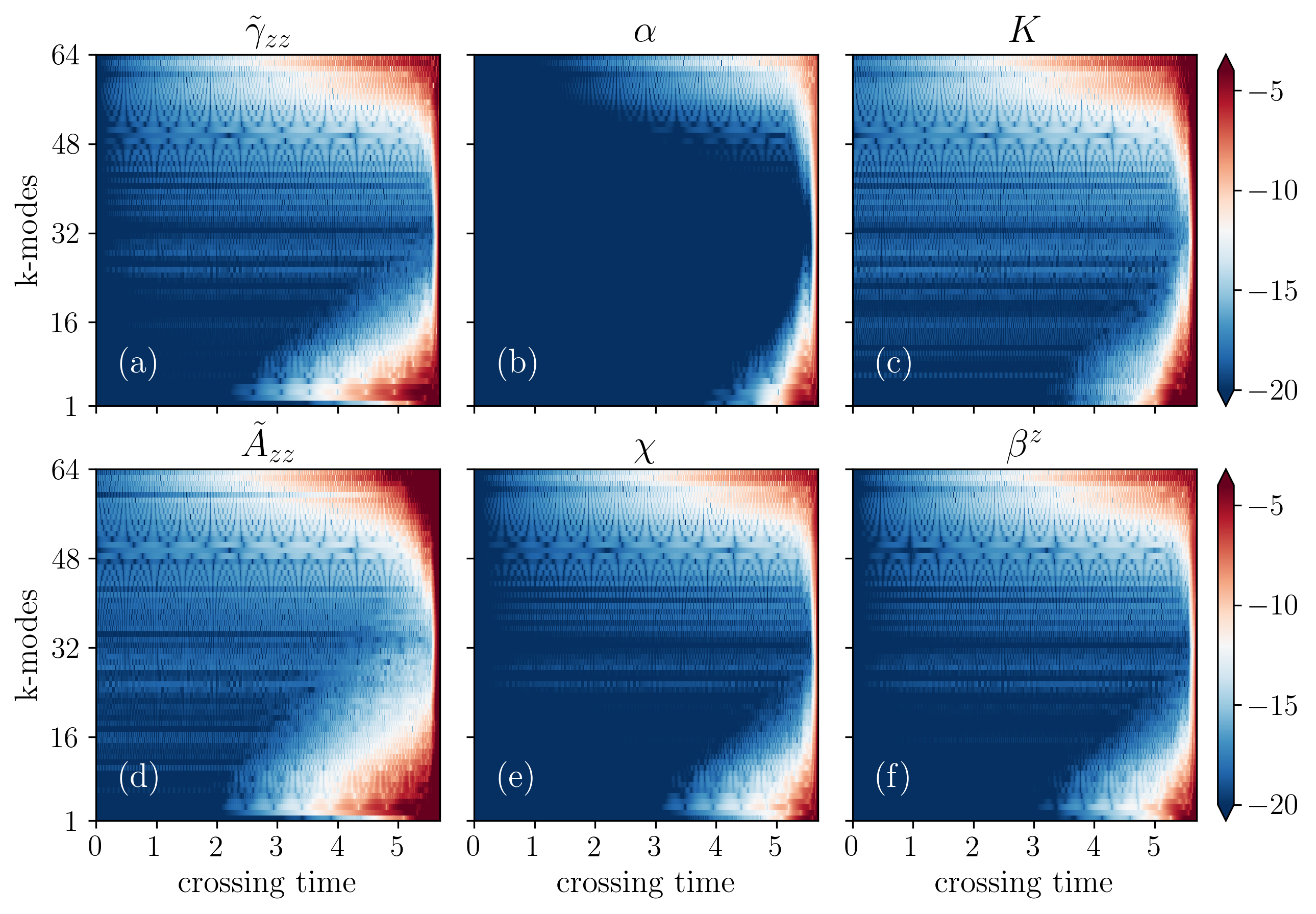}
\caption[...]
{\footnotesize{} The isotropic energy spectra {\it vs.} 
$k$ and $t$, for the robust stability test (RUN$_3$), without anti-aliasing filter. The code becomes unstable due to the aliasing error and the simulation crashes at $t\sim 5.7$. From $(a)$ to $(f)$, we show   $\widetilde{\gamma}_{zz}$, $\alpha$, $K$, $\widetilde{A}_{zz}$, $\chi$, and $\beta^z$.}
\label{robust6}
\end{figure}
\begin{figure}
\hspace{-5pt}
\includegraphics[height=40mm,width=119mm]{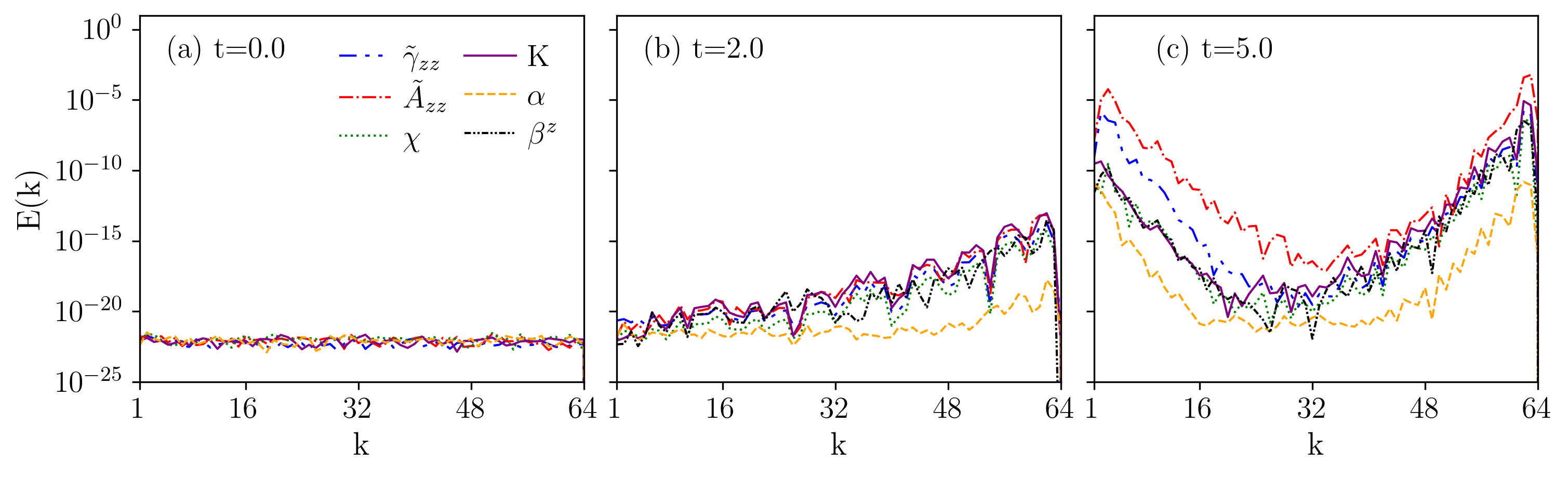}
\caption[...]
{\footnotesize{} Robust stability test for RUN$_3$. The energy spectra  are shown as a function of $k$, for $\widetilde{\gamma}_{zz}, \widetilde{A}_{zz}, \chi, K, \alpha$ and $\beta^z$, at 3 different times.}
\label{robust3}
\end{figure}

The second experiment consists of the robust stability test, a particularly stressful condition that is able to initiate numerical instabilities that exponentially grow in time.  The initial configuration consists of small random perturbations to the  Minkowski space, where $ds^2 = -dt^2 + dx^2 + dy^2 + dz^2$, where $\eta_{ij}$ is the unperturbed metric. The numerical noise is distributed among every BSSN dynamical field, let's say $f = f_0 + \varepsilon_f$, with $f_0$ being the unperturbed field. The perturbation is chosen to be $\varepsilon(f) \in [-10^{-10}, 10^{-10}]$. We chose a spatial domain $x, y, z \in [0,1]$, a number of grid points $N_z =128$, where the mesh size along $z$ is therefore $\Delta z= 1/128$ and the time-step is $\Delta t=10^{-3}$. To speed up calculations, we use only four grid points in the $x$ and $y$ directions. According to the literature, we use the harmonic slicing to evolve the lapse, namely $f(\alpha)=1$ in equation (\ref{slicing1}). We use the ``Gamma driver'' condition to evolve the shift \cite{cao2008reinvestigation}. For the first test (RUN$_3$) no anti-aliasing filter has been used, and the code crashes at $t \sim 5.7$ crossing times. In figure \ref{robust6} the energy spectra of the main BSSN  dynamical variables are reported as a function of time. From this figure is clear that aliasing instabilities, once again, come first from the extrinsic curvature (the trace and the traceless part) and then contaminate all the other dynamical fields, until the code crashes. In figure \ref{robust3}, we report the whole energy spectra at time $t=0$, 2 and 5, for the same fields.

\begin{figure}
\hspace{-10pt}
\includegraphics[height=80mm,width=119mm]{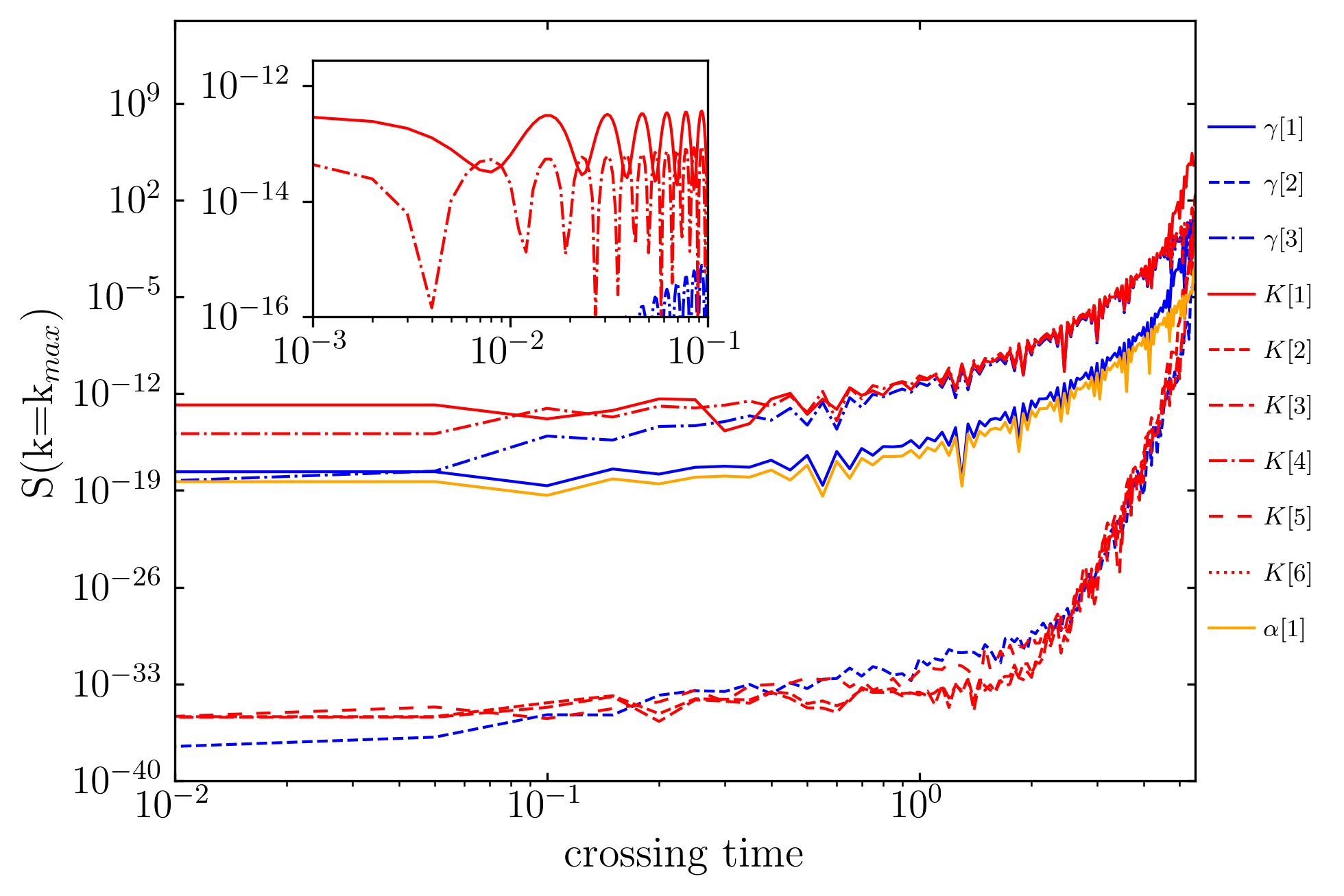}
\caption[...]
{\footnotesize{} 
Energy of the maximum mode, as a function of time, for the robust stability test with no anti-aliasing filter (RUN$_3$). In the inset, we show the initial stages, with the $K[1]$ term ($\propto R_{ij}$) being the dominant growing term. This term remains the most aliased, until the code crashes.}
\label{robustADM}
\end{figure}

The BSSN formalism involves a large number of auxiliary variables. It is instructive, in this regard, to monitor the more physical ADM fields, i.e. the metric $\gamma_{ij}$, the extrinsic curvature $K_{ij}$ and the lapse $\alpha$. In table \ref{tableADM}, we report in detail the various terms of the ADM evolution equations, together with the corresponding order of the products. The number of products varies from quadratic (as for example the time evolution of the metric), to very high nonlinearity (5$^{th}$ order products), as for the Ricci component of the extrinsic curvature. In order to inspect which is is the main candidate for the numerical instability, for each of these terms, we computed the energy spectra. Figure \ref{robustADM} shows the evolution in time of the $k=k_{max}$ Fourier mode, on a log-log scale. This analysis shows that the (non-physical) energy due to the aliasing phenomenon grows rapidly, until the code crashes. Different terms have different amplitudes and growth rates.

At this point, it is then useful to inspect the very initial stages, in order to reveal the precursor of such instability. The inset shows the earlier times of the simulation. As expected, the $K[1]$ term is the higher in energy since the initial time-steps, and remain always one of the most aliased during the simulation. Even at the end, this term keeps growing faster and reaching higher values. This behavior is due to the fact that $K[1]$ is proportional to the Ricci tensor $R_{ij}$ which involves more complex convolutions in the Fourier space, as discussed in Table \ref{table1}.

\begin{table}
\caption{\label{tableADM} Table of ADM terms, together with the corresponding order of the products. The highest nonlinearity is due to the $K[1]$ term, that is proportional to the Ricci tensor $R_{ij}$.  } 
{\def\arraystretch{1.4}\tabcolsep=4.0pt
\begin{tabular}{@{}lllll}
\hline\noalign{\smallskip}
             & term                     & \# products & $  $    \\
\noalign{\smallskip}\hline\noalign{\smallskip}
$\gamma \, [1]$  & $-2 \alpha K_{ij}$                 & 2    & $ $    
\\
$\gamma \, [2]$  & $\beta^k \partial_k \gamma_{ij}$   & 2    & $ $    
\\
$\gamma \, [3]$  & $\gamma_{ik} \partial_j \beta^k $ & 2    & $ $    
\\
$K \, [1]$  & $ \alpha R_{ij}  \sim \alpha [\gamma^{**} \partial_{*} \gamma_{**}][\gamma^{**} \partial_{*} \gamma_{**}]$                      & 5    & $ $    
\\
$K \, [2]$  & $ -2 \alpha K_{im} K^m_{~j} \sim \alpha K_{**} \gamma^{**} K_{**}  $          & 4    & $ $    
\\
$K \, [3]$  & $ -2 \alpha K K_{ij}  $                & 3    & $ $    
\\
$K \, [4]$  & $ -D_i D_j \alpha \sim [\gamma^{**} \partial_{*} \gamma_{**}] \partial_{*}\alpha $                   & 3    & $ $    
\\
$K \, [5]$  & $ \beta^k \partial_k K_{ij}$          & 2    & $ $    
\\
$K \, [6]$  & $ K_{ik} \partial_j \beta^k $          & 2    & $ $    
\\
$ \alpha $  & $ -\alpha^2 K  $                       & 3    & $ $
\\
\noalign{\smallskip}\hline
\end{tabular} }
\label{table1}
\end{table}

In order to show the benefit of the anti-aliasing filter and the robustness of the code, a second test (RUN$_4$) has been performed, by using our smooth filter at $k^* = N_z/3$, with the same parameters of RUN$_3$. The code remains stable for a long time and the Fourier spectra are well bounded (see for example \cite{Meringolo2021spectral}).

\subsection{Gowdy spacetimes}
We now analyze the aliasing instability via another stressful and unstable initial data, namely the Gowdy spacetime \cite{bona2010gowdy, Zlochower2005accurate}. The Gowdy wave test evolves a strongly curved spacetime, representing a vacuum cosmological model and therefore used to test codes in a strong field context. Such vacuum solutions describe an expanding/collapsing universe, and it contains plane-polarized gravitational waves. For this test, we will carry out our simulations in the collapsing direction, i.e. backward in time \cite{Alcubierre2003towards, New1998stable}.  The metric is given by
\begin{equation}
    ds^2 = -\frac{e^{\lambda /2}}{\sqrt{t}} dt^2 + \frac{e^{\lambda /2}}{\sqrt{t}} dz^2 + t e^\xi dx^2 + t e^{-\xi} dy^2,  
    \label{gowdym}
\end{equation}
which is clearly singular at $t=0$ (cosmological singularity).  The solutions are given by \cite{Babiuc2008implementation}
\begin{eqnarray}
    \nonumber
    \gamma_{xx} = t e^\xi~,~~~\gamma_{yy} = t e^{-\xi}~,~~~\gamma_{zz} = \frac{e^{\lambda/2}}{\sqrt{t}},
\end{eqnarray}
where  $\xi$ and $\lambda$ are functions of $z$ and $t$ only, and involves the Bessel functions \cite{Dumbser2018conformal}.

\begin{figure}[t]
\includegraphics[height=55mm,width=119mm]{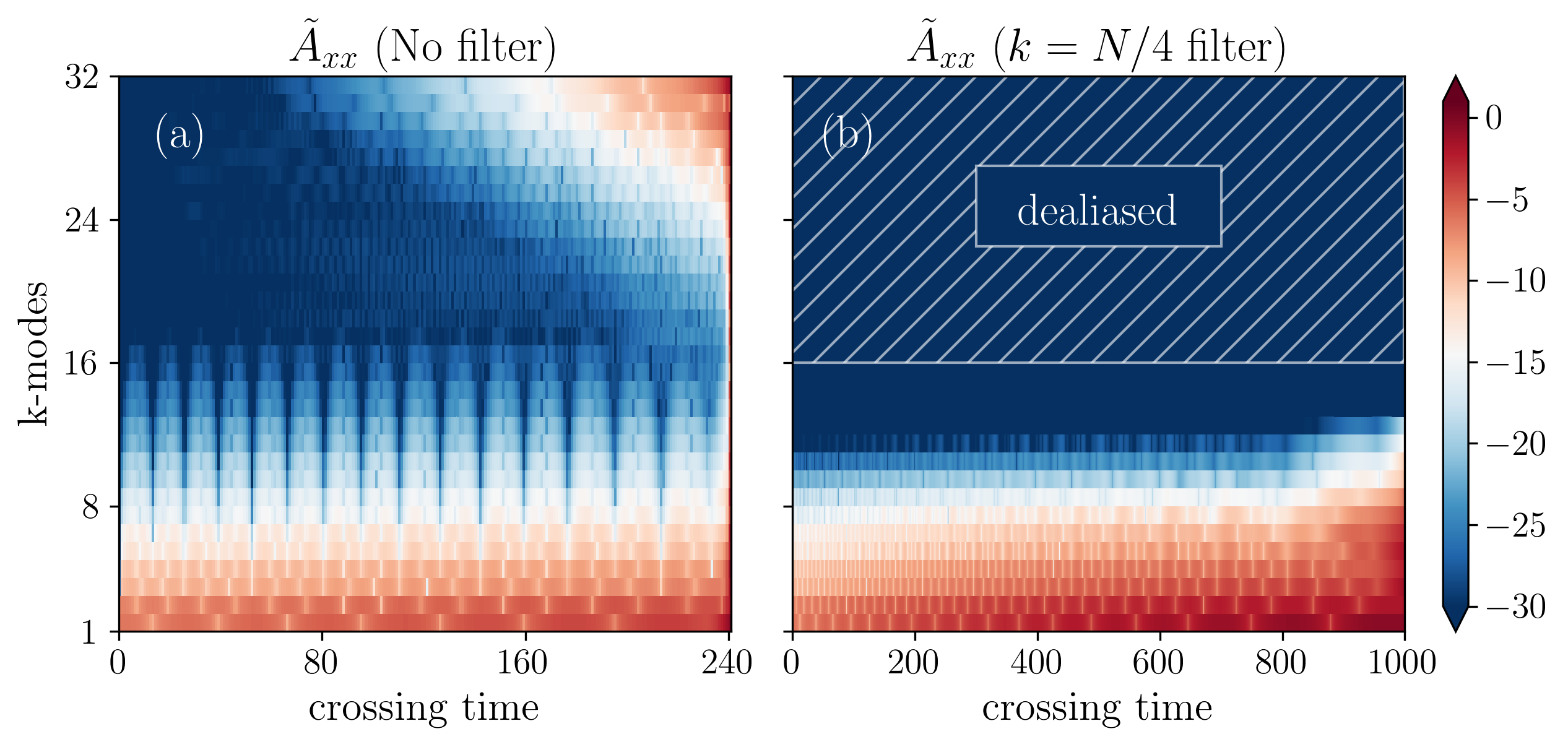}
\caption[...]
{\footnotesize{} Energy spectrum of $\widetilde{A}_{xx}$ as a function of time, for the Gowdy test (a) without anti-aliasing filter (RUN$_5$) and (b) with a $k^*=N_z/4$ filter (RUN$_6$).}
\label{gowdy_spec}
\end{figure}
To complete the initial data conditions, the extrinsic curvature is obtained as a time derivative of the metric. Since the metric (\ref{gowdym}) depends only on the $z-$direction, the spatial evolution is 1D. We chose a spatial domain $z \in [ 0, 1 ]$, with $N_z =64$, $\Delta t=5 \times 10^{-4}$ and the harmonic slicing. In the collapsing version, the time coordinate can be transformed such that the initial singularity is approached asymptotically, backward. The new time coordinate $\tau$ is defined by $\tau = c^{-1} \ln(t/k)$. In other terms, we replace $t \rightarrow k e^{c\tau}$, and the time-step $\Delta \tau$ was chosen to be negative. We choose a particular value of $t_0$ such that the initial slice is far from the cosmological singularity and, following the standard approach \cite{Daverio2018apples, Alcubierre2003towards}, we chose the twentieth zero of the Bessel function. This is an important challenge for a numerical code because there is both a small effect (the dynamics in $\gamma_{xx}$ and $\gamma_{yy}$) and a larger effect (dynamics in $\gamma_{zz}$) in the metric components.

As usual, the first run (RUN$_5$) is carried out without the anti-aliasing filter. The code becomes unstable and crashes at $t \sim 240$ crossing times. In panel (a) of figure \ref{gowdy_spec}, the time evolution of the spectrum of $\widetilde{A}_{xx}$ is shown. Note that with our choice of parameters, $k_{max}=N_z/2=32$.  The instabilities arise, again, from small wavelengths and propagate backward to larger scales. In order to stabilize the simulation, in the RUN$_6$, we set our smooth filter to $k^* = N_z /4$, together with the RSC condition. In this way the evolution becomes well-behaved until $t=1000$, as it can be seen from figure \ref{gowdy_spec} (right).

\begin{figure}[t]
\includegraphics[height=47mm,width=119mm]{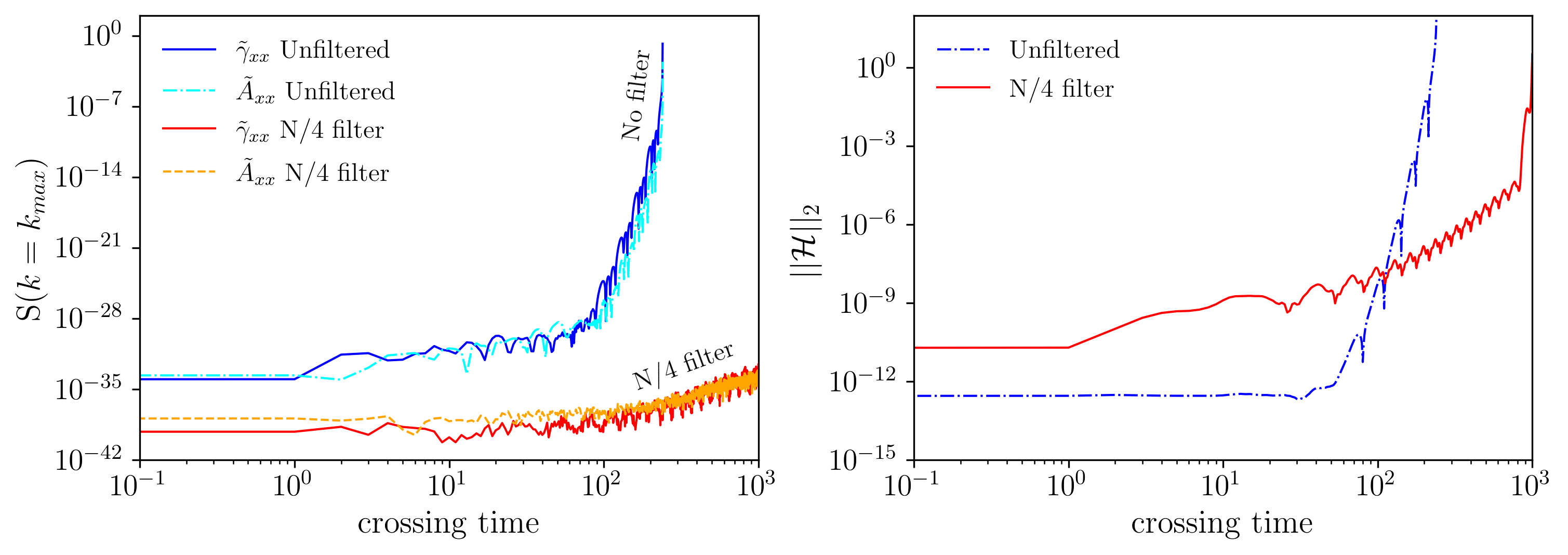}
\caption[...]
{\footnotesize{}Gowdy spacetime. Left: the evolution of the $k=32$ Fourier mode for the conformal metric $\widetilde{\gamma}_{xx}$ and the trace-free part of the extrinsic curvature $\widetilde{A}_{xx}$, with no filter (RUN$_5$) and with a smooth $k^*=N_z/4$ filter (RUN$_6$). Right: The evolution of the $L_2$ norm of the Hamiltonian, for both runs.}
\label{gowdy_2}
\end{figure}

In figure \ref{gowdy_2} we show the largest modes of $\widetilde{\gamma}_{xx}$ and $\widetilde{A}_{xx}$, for both the unfiltered (RUN$_5$) and the dealiased Gowdy test (RUN$_6$). To check the consistency of the results, we computed the
$L_2$ norm of the Hamiltonian constraints
\begin{equation}
    \Vert{\mathcal{H}}\Vert_2=\sqrt{\frac{\int_{\Omega} \mathcal{H}^2\, d\Omega\,\sqrt{|\gamma|}}{\int_{\Omega}d\Omega\,\sqrt{|\gamma|}}},
\label{norml2}
\end{equation}
where $d\Omega\,\sqrt{|\gamma|}$ is the volume element. This quantity, which should be null in the ADM representation of the Einstein field equations, is very small, as can be seen from figure \ref{gowdy_2} (right). However, the solution becomes suddenly unstable in the unfiltered case, while it keeps more reasonable low values in the filtered case, for longer times. In both cases, as expected in this kind of very difficult initial data, the solution inexorably blows up, in agreement with previous works \cite{Clough2015grchombo}. Overall, as it can be seen, also in this more challenging test our pseudo-spectral code is able to handle the numerical evolution and is stable for the full 1000 crossing times, with violation errors that are comparable (or smaller) than in previous works \cite{Alcubierre2003towards}.

%
%


\subsection{Head-on collision}
\label{Headon}

As a final test, we present results about the head-on collisions of two equal-mass Misner-Wheeler-Brill-Lindquist (MWBL) black holes \cite{Brill1963interaction, Misner1957classical}. The MWBL data represent conformally flat slices of multiple black hole space-times with $n$ punctures. In this formalism, each black hole is parametrized by $m_j$ and ${\bf r}_j$, namely the mass parameter and the position of the $j-th$ black hole, respectively. For the head-on test, the ADM mass is given by the sum of $m_j$. Both the ADM linear and angular momentum are set to zero, this means that the black holes start without boost and spin.

The extrinsic curvature is set initially to zero, with $ \widetilde{A}_{ij} = K = 0$, and with a conformal factor that is
\begin{equation}
\nonumber
   \chi = \left[ 1 + \frac{m_1}{2 |{\bf r} - {\bf r}_1|} + \frac{m_2}{2 |{\bf r} - {\bf r}_2|} \right]^{-4}.
\end{equation}
In this case the 3D computational domain extends over $x, y, z \in [0,25]$
We set the mass parameters and the initial location of the $j-th$ puncture to be $m_j = 0.5$ and $\bf{C}_j = $ $\{ 12.5,12.5 \pm2 ,12.5 \}$, respectively. As suggested by \cite{Dumbser2018conformal}, we first define $r^* \stackrel{d}{=} \left( r^4 + 10^{-24}\right)^{\frac{1}{4}}$, where $r$ is the coordinate distance of a grid point from the puncture. The lapse is then initially set to
\begin{equation}
\nonumber
    \alpha = \frac{1}{2} \left[ \frac{1- \frac{1}{2} (m_1 / r_1^*) - \frac{1}{2}  (m_2 / r_2^*)}{1+ \frac{1}{2} (m_1 / r_1^*) + \frac{1}{2} (m_2 / r_2^*)} +1 \right],
\end{equation}
while the shift is initially null. We use the harmonic slicing to evolve the lapse, and the ''Gamma driver'' condition to evolve the shift \cite{cao2008reinvestigation}.

\begin{figure}[t]
\includegraphics[height=50mm,width=119mm]{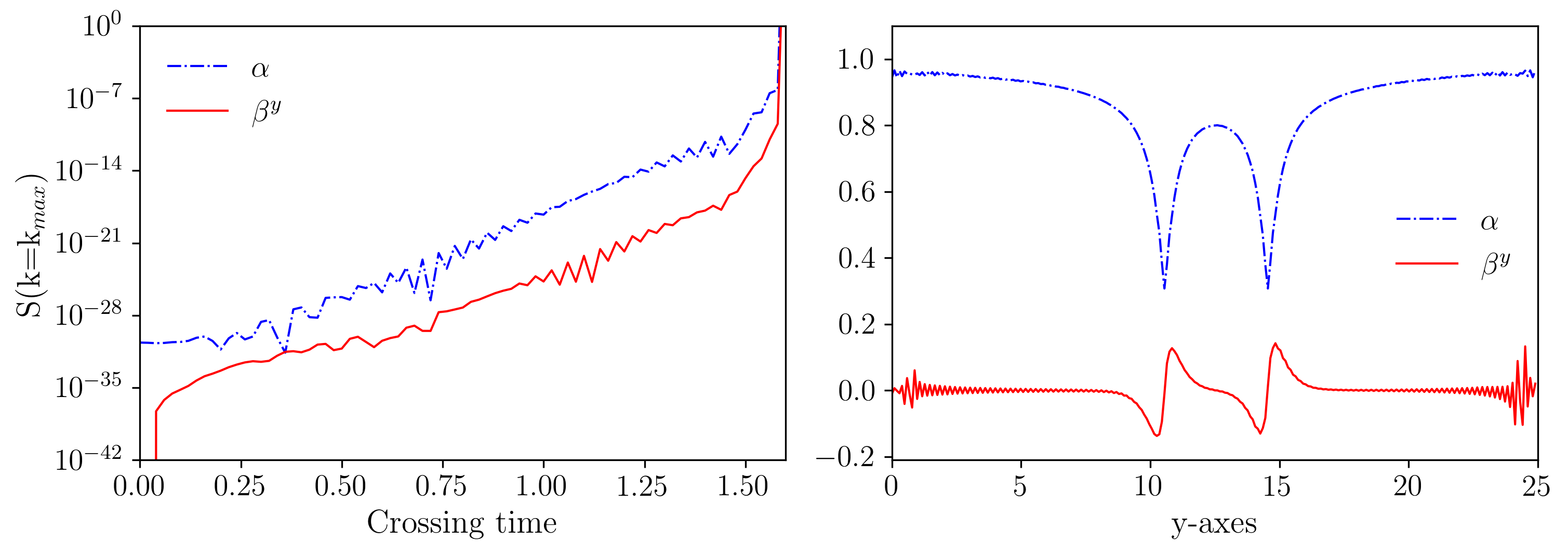}
\caption[...]
{\footnotesize{} The head-on collision test (RUN$_7$). Left: the time-growth of the mode $k = k_{max}$, for both the lapse and the shift. The code crashes at $t\sim 1.6$. Right: 1D section in the $y-$direction for both fields, at $t=1.5$, just before the code crashes. Note that the instabilities arise from the boundaries and propagates into the domain. }
\label{head1}
\end{figure}

For the first run, we stress the code without an anti-aliasing filter  (RUN$_7$). As expected, such a heavy test quickly induces aliasing instabilities and the code suddenly crashes. Since this is a simulation with a 3D mesh, with $N=256$ grid points for each Cartesian direction, the maximum mode is at about $N_k = \frac{1}{2} \sqrt{3 N^2 } \sim 222$. In figure \ref{head1} (left) we show the evolution of the $k=k_{max}$ mode for the lapse $\alpha$ and the shift $\beta^y$. In the right panel, a 1D cut of the same fields is shown, at $t=1.5$, just before the code blows up. For the symmetry of the problem, the cut is along the $y$-axis, at $N_x/2$ and $N_z/2$ . One can see the growth of ripples at the boundary regions, due to the propagation of internal waves and their interference related to periodicity. These non-physical fluctuations grow up exponentially in time and lead to unstable solutions. A second simulation (RUN$_{8}$) has been carried out by using a smooth filter with $k^*=N/2$. The filter surely improves the stability of the simulation, but the code crashes later than the RUN$_7$, namely at $t \sim 4.2$. This indicates that the boundary might be now the cause of the aliasing problems.

\begin{figure}[t]
\includegraphics[height=77mm,width=119mm]{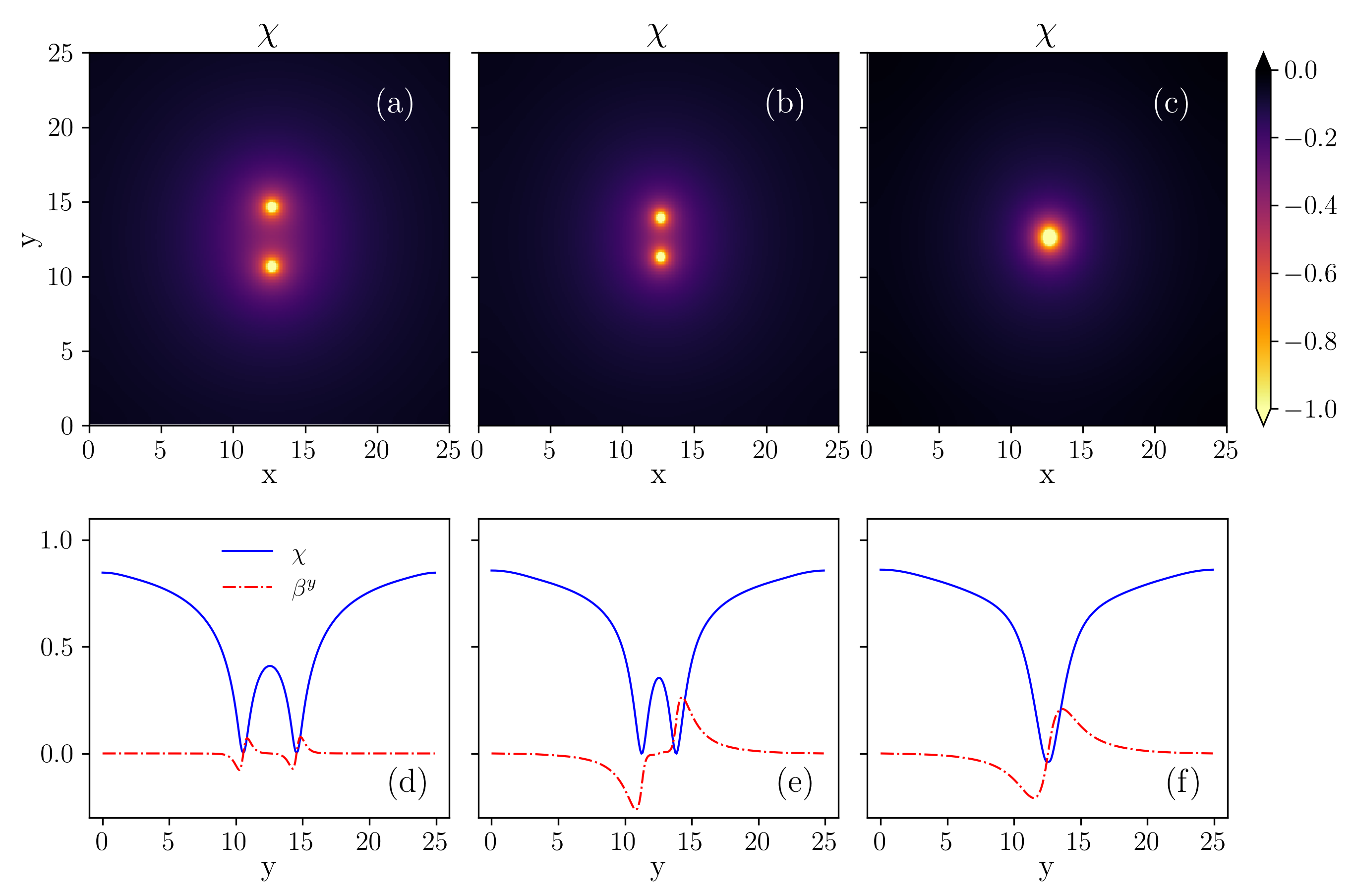}
\caption[...]
{\footnotesize{} Head-on collision with a stabilizing procedure (RUN$_8$). A smooth filter with $k^*=N/2$ and the IHB has been used. Top row: $xy-$plane at time $t=1, t=13$ and $t=26$ for the conformal factor $\chi$. Bottom row: 1D cut in the $y-$direction for $\chi$ and $\beta^y$, at same times. The field remains well-behaved at the boundaries. The merger occurs at $t \sim 21$ and the simulation is carried out until $t=50$.}
\label{head6}
\end{figure}

\begin{figure}[t]
\centering
\includegraphics[height=70mm,width=100mm]{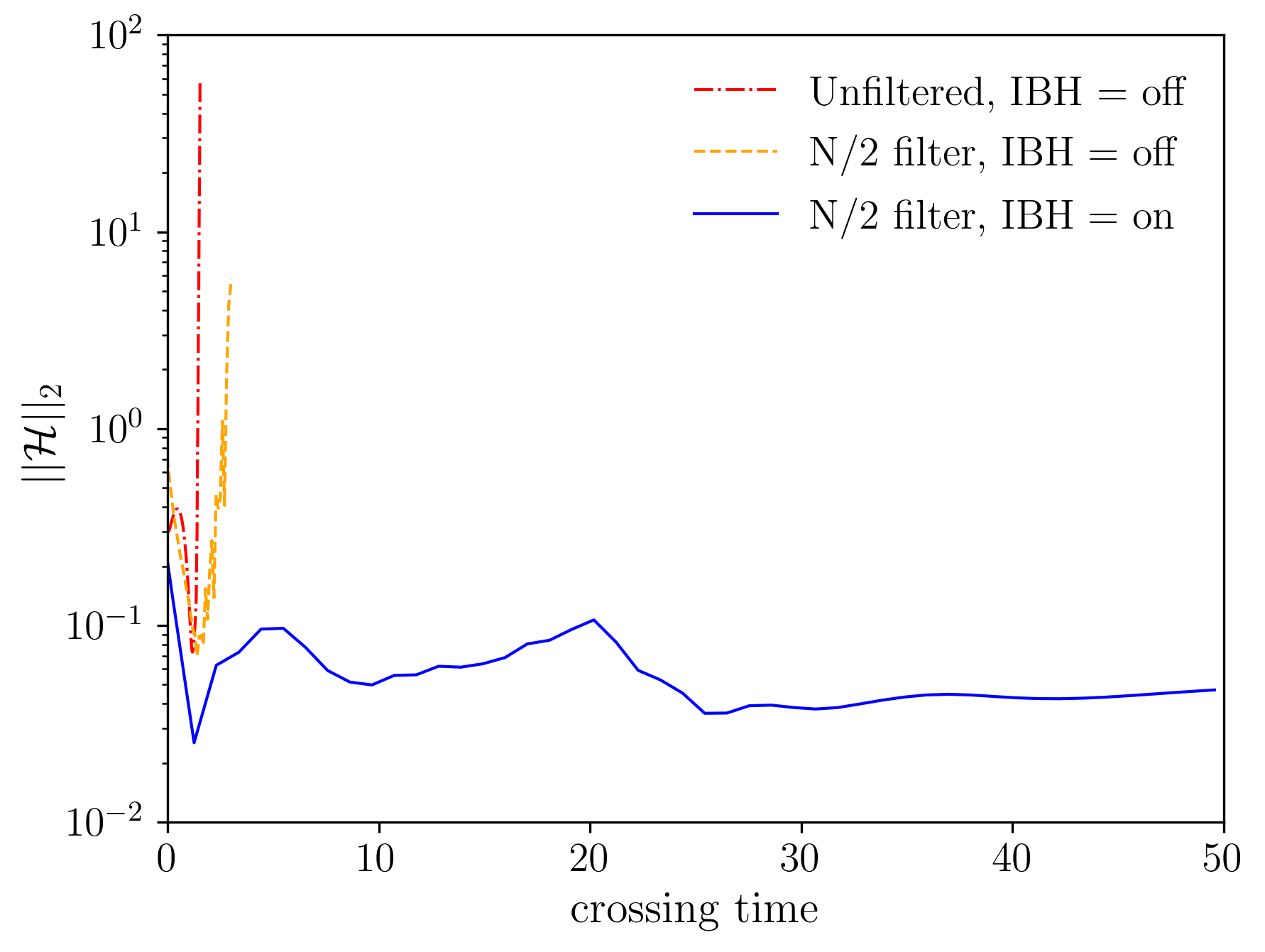}
\caption[...]
{\footnotesize{} $L_2$ norm of the Hamiltonian constraint, for the head-on collision of two equal masses black holes. The merger occurs at $t\sim 21$.
The solid line represents the test with IHB on and a smooth filter with $k^* = N/2$ (RUN$_9$), while the dashed (RUN$_7$) and the dot-dashed (RUN$_8$) represent the unstable runs.}
\label{H2head}
\end{figure}

In order to definitively stabilize the code, we carried out a third simulation (RUN$_9$) with the viscous approach, switching on the IHB technique described in \cite{Meringolo2021spectral}. We carried out the simulation several times after the merging of the two black holes. In the top row of figure \ref{head6} is reported a section in the $xy-$plane (at $N_z/2$) of the conformal factor $\chi$, at three times, namely $t=1, 13$ and $26$. The merging occurs at $t\sim 21$, even though the simulation is carried out until $t=50$. In the bottom-row, we show, at the same times, 1D cuts in the $y-$direction (at the middle of the lattice) of $\chi$ and $\beta^y$. The numerical artifacts in figure \ref{head1} have been suppressed thanks to the combination of filtering and IHB technique. For completeness, we monitor the evolution in time of all the BSSN constraints during the simulation, for the three head-on collision tests. In figure \ref{H2head} we report the $L_2$ norm of the Hamiltonian constraint, described by Eq.~(\ref{norml2}), for the RUN$_7$--RUN$_9$.  As it can be seen, the run with the hyperviscous boundaries is stable long after the collision, manifesting also a lower violation of the constraints.

The SFINGE code, with the above filtering and boundary treatments, is able to handle such difficult gravitational dynamics. A similar strategy can be used for a variety of studies, including the inspiring binaries and the multiple black holes systems \cite{Pretorius2006simulation, Matzner1998initial, Lousto2008foundations}, which will be presented in future works.

\section{Discussion}
\label{discuss}
We have presented a spectral analysis of the numerical instabilities that usually affect numerical relativity, by using standard numerical testbeds. Our approach is based on the BSSN formalism in vacuum conditions. We have studied different initial data, namely the gauge wave test, the robust stability test, the Gowdy waves test, and the head-on collision of two equal-masses black holes.

Our numerical method is based on a pseudo-spectral technique, where we compute spatial derivatives via simple Cartesian FFTs. Before the numerical simulations, we discuss a brief overview of the BSSN equations and we presented an analysis of the aliasing instabilities. We studied the arising of such instabilities by monitoring the Fourier spectra of the dynamical variables, showing how the numerical noise due to the aliasing effects affects the highest Fourier modes, propagating eventually to the whole spectrum. These growing modes proliferate over all the fields.

We presented a strategy that mitigates such instabilities. First, we have discussed a smooth anti-aliasing filter that is able to suppress numerical instabilities and improve the simulations. In order to show the goodness of the numerical simulations, in the gauge wave test we have matched the numerical and the analytical solution, with and without the filter, with an excellent agreement for the dealiased run. For the other tests, we have monitored the highest Fourier modes and the Hamiltonian violation, showing a net improvement of the simulation by using our smooth filter.

For inhomogeneous metrics, as in the case of the head-on collision of two black holes, the SFINGE code relies on a novel technique, namely the implicit hyperviscous boundary condition. This consists of matching of the ideal solution in the center of the domain, based on the pseudo-spectral solution of the BSSN system of equations, with a semi-implicit, second-order Crank-Nicholson technique at the boundaries, where we added hyper-viscous diffusion. The strategy is able to suppress both spurious boundary effects and aliasing phenomena.

The analysis confirms that spectral codes are very accurate for numerical relativity, suggesting however that great care must be taken about the numerical computation of highly nonlinear terms, where the aliasing effect is more pronounced. These aliasing instabilities might be mitigated via the use of spectral filters, which are in principle easy to implement in any spectral representation. Future works will concentrate to extend the present algorithm to the dynamics of inspiraling binary systems, as well as to the dynamics of many-body problems. Finally, we plan to export the method to the solution of the gravitational fields in presence of matter as well as its coupling with electromagnetic fields, in the framework of general relativistic magnetohydrodynamics.


\begin{acknowledgements}
The simulations have been performed at the Newton HPPC Computing Facility, at the University of Calabria. 
\end{acknowledgements}




\end{document}